\documentstyle[11pt, aasms4]{article}
\input epsf
\tighten
\begin{document}
\title{Measuring cosmological bulk flows via the kinematic Sunyaev-Zeldovich 
effect in the upcoming cosmic microwave background maps.}

\author{A. Kashlinsky}
\affil{Raytheon ITSS\\ Code 685, Goddard Space Flight Center, Greenbelt,
MD 20771\\
	e--mail: kashlinsky@stars.gsfc.nasa.gov}

\author{F. Atrio--Barandela} 
\affil{F{\'\i}sica Te\'orica. Facultad de Ciencias.\\
	Universidad de Salamanca, 37008 Spain.\\
	e--mail: atrio@orion.usal.es}

\begin{abstract}
We propose a new method to measure the possible large-scale bulk flows in the 
Universe from the cosmic microwave background (CMB) maps from the upcoming 
missions, MAP and Planck. This can be done by studying the statistical 
properties of the CMB temperature field at many X-ray cluster positions. At 
each 
cluster position, the CMB temperature fluctuation will be a combination of the 
Sunyaev-Zeldovich (SZ) 
kinematic and thermal components, the cosmological fluctuations and the 
instrument noise term. When averaged over many such clusters the last three 
will 
integrate down, whereas the first one will be dominated by a possible bulk flow 
component. In particular, we propose to use all-sky X-ray cluster catalogs that 
should (or could) be available soon from X-ray satellites, and then to evaluate 
the dipole component of the CMB field at the cluster positions. We show that 
for 
the MAP and Planck mission parameters the dominant contributions to the dipole 
will be from the terms due to the SZ kinematic effect produced by the bulk flow 
(the signal we seek) and the instrument noise (the noise in our signal). 
Computing then the expected signal-to-noise ratio for such measurement, we get 
that at the 95\% confidence level the bulk flows on scales $\geq 100h^{-1}$Mpc 
can be probed down to the amplitude of $< 200$ km/sec with the MAP data and 
down 
to only $\simeq 30$ km/sec with the Planck mission.
\end{abstract}

\keywords{Cosmology: Cosmic Microwave Background. Large Scale Structure. 
Galaxies: Clusters. }

\newpage 
{1. \bf Introduction.}\\
Peculiar motions trace the
overall mass distribution and it is important
to determine the coherence scale and amplitude of bulk flows. Current 
measurements range from bulk flows as high as 700 km/sec on scale of 
$\sim$150$h^{-1}$Mpc (Lauer \& Postman 1994) to finding little peculiar
motion on scales beyond $\sim$70$h^{-1}$Mpc (see Willick 
2000 for review). It is thus important
to find alternative ways to test for such bulk flows.
One alternative way to measure peculiar flows 
is via the kinematic Sunyaev Zeldovich (SZ) effect produced on the cosmic 
microwave background (CMB) photons from
the hot X-ray emitting gas in clusters of galaxies (Zeldovich \& Sunyaev 
1969). Such program is already being 
undertaken in the SuZIE project which plans to measure motion of 40
clusters at $z$=0.1-0.3 and determine the peculiar velocity of each 
cluster to a precision of 700 km/sec.
 
In this {\it letter} we propose to use CMB data from the MAP and 
Planck missions to measure the bulk flows in a quick, cheap and 
efficient way. X-ray cluster catalogs to be available shortly based on ROSAT, 
ASCA and XMM measurements, will 
provide locations of the SZ sources on the CMB sky and a 
reasonable estimate of the cluster electron density distribution and 
temperature. If there are significant bulk motions they will 
leave an imprint via the cumulative kinematic SZ 
effect and can be uncovered by cross-correlating the temperature field at the 
cluster positions in the MAP and Planck CMB maps. Such bulk motions would 
produce a significant dipole component in the temperature field evaluated at 
the cluster locations. By averaging temperature fields at enough cluster 
positions, the thermal SZ and other noise components will integrate down 
enough to reveal possible bulk motions out to $\sim$ 300-400 km/s on 
scales $\geq$50-100$h^{-1}$Mpc. 

{2. \bf Dipole of the cumulative SZ kinematic effect}.\\
Consider the CMB field at a beam centered on one isothermal X-ray emitting 
cluster at 
the angular position $\vec y$. If the cluster is moving with the line-of-sight 
velocity $v_r$ with respect to the CMB rest 
frame, the SZ CMB fluctuation at frequency $\nu$ at 
this position will be $\delta_\nu(\vec y)$=$\delta_{th}(\vec y)G(\nu)$+$ 
\delta_{kin}(\vec y)H(\nu)$, with $
\delta_{th}$=$\tau T_{\rm virial}/T_{e,ann}$ and
$\delta_{kin}$=$\tau v_r/c$
(e.g. Phillips 1995). Here $\tau$ is the projected 
optical depth due to Compton 
scattering, $T_{\rm virial}$ is the cluster virial temperature and 
$k_BT_{e,ann}$=0.5 MeV. (For expressions for the spectral dependence 
of the two SZ components, $G(\nu), H(\nu)$ see e.g. Birkinshaw (1999).) 
Normally the thermal term 
dominates for individual clusters, but if averaged over many clusters moving 
at a significant bulk flow with 
respect to the CMB rest-frame, the former will integrate down 
$\propto 1/N_{\rm cluster}$, while the kinematic term will reflect coherent 
motions with amplitude $V_{\rm bulk}$.

In order to minimize the contribution from other sources, we will start with 
CMB maps from which the cosmological dipole component was subtracted down to 
$\sigma_d$. This uncertainty is already $\sigma_d=7\mu$K 
(Fixsen et al. 1996) and should be significantly smaller in MAP observations.
The MAP radiometers produce a raw temperature measurement that is the 
difference between two points on the sky $\sim 140^o$ apart
which introduces an error on the dipole determination, due to correlated
noise, of the order of $0.1\mu$K; the dominant uncertainty in the 
dipole will be due to confusion from the galactic foreground
(G. Hinshaw, private communication). 
After the cosmological CMB dipole 
subtraction the CMB fluctuation in band $\nu$ at position $\vec 
y$ centered on a known X-ray cluster will be $
\delta_\nu(\vec y)$=$\delta_{th}(\vec y)
G(\nu)$+$[\delta_{kin}(\vec y)$+$\delta_{\rm CMB}(\vec y)]H(\nu)$+$r(\nu)$. 
Here $r(\nu)$ is the instrument noise at frequency $\nu$ and $\delta_{\rm 
CMB}(\vec y)$ is the cosmological CMB component whose dipole is now 
$\sigma_d^2$.
Consider the dipole component of $\delta_\nu(\vec y)$  with the dipole 
amplitude $C_1$ normalized so that a coherent motion at velocity $V_{\rm 
bulk}$ would lead to the dipole amplitude of $V_{\rm bulk}^2/c^2$. 
When computed from the total of $N_{\rm cluster}$ positions the dipole of the 
noise term becomes $\langle r^2(\nu) \rangle/N_{\rm cluster}$. 
The cosmological signal gives rise to two different dipole contributions:
1) the cosmological dipole has not been perfectly removed so 
the temperature anisotropies at the cluster locations 
sample the residual dipole $\sigma_d$; and 2) even if
all the cosmological dipole had been removed the 
intrinsic CMB temperature anisotropies
could be seen as an extra dipole noise source. The latter 
contribute  $\sigma_{\rm CMB}^2/N_{\rm cluster}$, with
$\sigma_{\rm CMB}$ being the variance of the cosmological temperature
field on 
the smallest angular scales probed by the experiment.
Thus for $N_{\rm cluster}\gg1$ the dipole of (3) becomes:
\begin{equation}
C_{1,\nu} \simeq C_{1,kin}H^2(\nu) +C_{1,th} G^2(\nu) + [
\sigma^2_{\rm CMB}/ N_{\rm cluster} + \sigma^2_d] H^2(\nu) 
+ \langle r^2(\nu) \rangle/N_{\rm cluster}
\label{C1} 
\end{equation}
where it was assumed that for each individual X-ray cluster the thermal SZ 
term dominates. 

{3. \bf Signal and noise terms.}\\
We now estimate the amplitude of the signal, $C_{1,kin}$, and noise
terms in eq. 
(\ref{C1}).

{3.1. \bf Kinematic component.}\\
Assuming that cluster properties are independent of their velocities, this 
term is $C_{1,kin}$=$T_0^2 \langle \tau_i\rangle^2 \frac{V_{\rm 
bulk}^2}{c^2}$.  $T_0$ is the CMB temperature, and 
$i$ refers to the frequency band. The effective optical depth is 
affected by the beam dilution. To evaluate the expected mean optical depth 
accounting for the beam dilution effects we proceed as follows: for X-ray 
clusters the electron density profile can be approximated by the $\beta$-model. 
For isothermal and spherical X-ray gas distribution, 
the optical depth as a function of the angular 
distance from the cluster center would be $\tau = \tau_0 (1+\theta^2 
D^2/r_{\rm core}^2)^{-\frac{3\beta-1}{2}}$. 
Here $D$ is the distance to the cluster and $r_{\rm core}$ is its core radius.
For each individual cluster observed in a CMB search, this expression needs to 
be convolved with the beam profile. The effective optical depth becomes
$\tau_i \equiv \tau_0 \Psi_i(D)$, where $\Psi_i$ accounts for the beam 
dilution effects. To evaluate $\Psi_i(D)$ we compiled a list of 37 X-ray 
clusters 
with measured $\beta, r_{\rm core}$ and 2.4 KeV$<$$T$$<$14.6 KeV from Arnaud \& 
Evrard (1999), Myers et al. (1997) and Neumann \& Arnaud (1999). We then 
computed 
$\Psi_i(D)$ for each individual cluster 
and the mean and the r.m.s. in $\Psi_i(D)$ evaluated over the ensemble 
of clusters. Fig.1 plots the mean and the r.m.s. values 
of $\Psi_i(D)$ vs $D$ for the largest and smallest MAP 
beams. The figure shows that for the purposes of estimating the 
magnitude of the kinematic dipole component we can assume that a ``universal" 
profile for the cluster optical depth exists to within an uncertainty of 
$\sim$10-20\%. 

The value of $\langle \tau_0\rangle$ can be estimated from the observed 
cluster properties and their X-ray luminosity function. Cooray (1999 - Table 
1) compiled a list of 14 X-ray clusters with measured SZ thermal components, 
$\Delta T_{\rm SZ}$, with the X-ray luminosity in the [2-10] KeV range, 
$L_X$(2-10 KeV), between 1.6$\times$$10^{43}$ and 3.6$\times$$10^{45} 
h^{-2}$erg/sec. 
A linear regression fit to the data gives $\tau_0$=(4.8$\pm$1.0)$\times$$ 
10^{-3} [L_X$(2-10KeV)$/10^{44}h^{-2}{\rm erg/sec}]^{\alpha_\tau}$ with 
$\alpha_\tau$=0.41$\pm$0.12.  Note that for isothermal X-ray clusters emitting 
due to thermal 
Bremsstrahlung, $L_X \propto n_e^2 T_{\rm virial}^{1/2}$, and obeying the 
observed X-ray luminosity - temperature relation, $L_X$$\propto$$T_{\rm 
virial}^\gamma$ with $\gamma$$\simeq$2.5 (Mushotzky \& Scharf 1997, Allen \& 
Fabian 1998, Arnaud \& Evrard 1998),  one expects $\tau$$\propto$$
L_X^{0.4}$ (e.g. Haenhelt \& Tegmark 1996). These relations show little 
evolution out to $z$$\sim$0.5 (Mushotzky \& Scharf 1997; Schindler 
1999) and are valid for cluster catalogs of depth $<$200$h^{-1}$Mpc 
needed for this project.

The mean optical depth can now be computed from measurements of the 
X-ray luminosity function (XLF). XLF has now been determined very accurately 
from the ROSAT BCS sample out to $z$$\leq$0.3 (Ebeling et al., 1997). The 
sample is 90\% complete for fluxes $\geq$4.45$ \times$$10^{-12}$ 
erg/cm$^2$/sec in [0.1-2.4] KeV band. The XLF is of the Schechter type $n(L_X) 
dL_X$=$ n_* (L_X/L_*)^{\alpha_X} \exp(-L_X/L_*)dL_X$ with $\alpha_X $$\simeq 
$$-1.8$ and bolometric $L_*$$\simeq $$9.3$$ \times$$ 10^{44}h^{-2}$erg/sec or 
$L_*$(2-10 KeV)$\simeq$$ 3.2$$ \times$$ 10^{44}h^{-2}$erg/sec. For these XLF 
parameters and $V_{\rm bulk}$=600 km/sec and a constant lower limit on the 
absolute X-ray luminosity $L_0$$\simeq$$ 
5$$\times$$ 10^{-3}L_*$, corresponding to the [0.1-2.4]KeV flux $4.5$$\times 
10^{-12}$ erg/cm$^2$/sec at 50$h^{-1}$Mpc, we get $\sqrt{C_{1,th}} $$\sim$$ 9 
\mu$K dropping to $\sim $$6 \mu$K for $L_0$$\simeq$$ 2$$\times$$ 10^{-3}L_*$. 

{3.2. \bf Thermal component}\\
The residual dipole from the SZ thermal component comes from the 
finite number of Poisson-distributed X-ray clusters and would
integrate $\propto 1/N_{\rm cluster}$. Since cluster 
properties are independent of position, the dipole contribution is 
$C_{1,th}$$\simeq$$\langle (\Delta T_{SZ}) \rangle^2  {\cal D}^2_{\rm 
cluster}$.
Here $\langle (\Delta T_{SZ}) \rangle$ is the mean amplitude of the thermal SZ 
temperature fluctuation produced by the observed X-ray clusters and 
${\cal D}_{\rm cluster}$=$\langle \cos\theta\rangle$, with $\theta$ being the 
cluster azimuthal angle, is the mean dipole of the cluster distribution out to 
the depth on which the bulk motions are probed. Assuming the scaling of 
$\tau$ vs $L_X$ above 
and integrating it over $L_X$$\geq$$L_0$=const gives $\langle 
(\Delta T_{SZ}) \rangle$=12$\mu$K for $L_0$=$0.002L_*$ and 
$\langle (\Delta T_{SZ}) \rangle$=20$\mu$K 
for $L_0$=$0.005L_*$. 
We used the Abell/ACO catalog (Abell 1958, Abell, Corwin \& Olowin 1989) to 
verify that the 
angular distribution of clusters has ${\cal D}^2_{\rm 
cluster}$$\propto$$N^{-1}_{\rm cluster}$. Statistically, one
expects ${\cal D}^2_{\rm cluster} 
N_{\rm cluster}$=1/3. For the ACO catalogue out to 
$200h^{-1}$Mpc we get ${\cal D}^2_{\rm cluster} N_{\rm cluster}$$\sim$0.2-0.3. 
The numbers are consistent with related dipole and monopole 
parameters of the gravitational field due to clusters ($D_S, M_S$=$\sum
r^{-2} \cos\theta, \sum r^{-2}$) from Fig.1 of Scaramella et al. 
(1991), when corrected for the fact that $D_S,M_S$ are dominated by more nearby 
clusters.
The different spectral dependence, $G(\nu)$, of thermal SZ can be used to 
reduce 
this term further. Thus the contribution 
of this term to the dipole noise term in eq. (\ref{C1}) would be 
$\sqrt{C_{1,th}}< 10 N^{-1/2}_{\rm cluster} \mu$K. (The thermal component 
dipole can also be seen to be small from extrapolation to $l$=1 from Figs. 1,2 
in theoretical computation of the thermal SZ power spectrum by Atrio-Barandela 
\& M\"ucket 1999, or Fig. 8 in Refreiger et al. 2000). 

{3.3. \bf Dipole cosmological components.}\\
There will be two independent contributions to the dipole noise from         
cosmological terms: from the residual dipole uncertainty and  
from the CMB fluctuations leaving a residual dipole when 
evaluated over a finite number of points on the sky $(N_{\rm cluster})$.

The first contribution will come from the cosmological dipole which can be 
eliminated from the CMB maps down to the 68\% uncertainty of 
$\sigma_d$ (=7$\mu$K for FIRAS). 
With $N_{\rm cluster}$ we will approach this 
systematic component as $\sigma_d^2 [1+{\rm O}(N_{\rm cluster}^{-1})]$.  
Because the correlation angle of the temperature anisotropies
is larger than the pixel size,
this component can be further decreased using a low-pass filter by 
performing the analysis on $\delta(\vec y_{\rm cluster})$--$\delta(
\vec y_{\rm ngb})$, where $\vec y_{\rm ngb}$ is the neighboring pixel 
that does not contain another cluster and is $\Delta \theta_{\rm ngb}$
away from the original pixel.
The dipole noise will then be reduced to 
$\sigma_d \; (\Delta \theta_{\rm ngb}/180^{\rm o})$,
while the noise variance will increase by only $1/N_{\rm ngb}$.
The second contribution will come from the cosmological temperature anisotropy 
at each cluster location. The r.m.s. temperature anisotropy $\sigma_{\rm CMB}$
on the smallest scales probed by MAP is model 
dependent, but could be of the same order of magnitude as the pixel noise 
or even larger and it has the same frequency 
dependence as the kinematic SZ effect. 
Its contribution is $\sigma_{\rm CMB}^2/N_{\rm cluster}$, and can be 
reduced by the low-pass filtering discussed above. If necessary, 
$\sigma_{\rm CMB}$ can be reduced further with Wiener filtering (cf. Haenhelt 
\& 
Tegmark 1996) designed to minimize the difference between the filtered
signal, ${\tilde \delta}_{\rm CMB}$, and the \underline{instrument noise}, $r$.
We computed the residual (Wiener-filtered) variance for two cold-dark-matter 
(CDM) models, the standard CDM with $\Omega$=1, and the cosmological constant 
dominated CDM with $\Omega$=0.3 with $\sigma_{\rm CMB}$=(57--93)$\mu$K in the 
MAP bands. After the filtering with the noise
multipoles of MAP the numbers reduce to $<$$\frac{1}{2}$$\langle r^2
\rangle^{1/2}$; this component adds in quadrature to the instrument noise.

{3.4. \bf Instrument noise component}.\\
If the instrument noise in the given band is $r(\nu)$, its 
contribution to the dipole term in (\ref{C1}) is $\langle r^2(\nu) 
\rangle /N^{-1}_{\rm cluster}$. The MAP mission (http://map.gsfc.nasa.gov) has 
5 bands from 22 to 90 GHz and the r.m.s. noise at the end of two years should 
reach $\langle r^2(\nu) \rangle^{1/2}$=35$\mu$K per 0.3$\times$0.3 deg 
pixel (the MAP beams range from 0.93 to 0.2 deg and its total lifetime would be 
at least 27 months). The PLANCK satellite 
(http://astro.estec.esa.nl/Planck) has Low and High Frequency Instruments (LFI 
and HFI). The LFI has four bands from 30 to 100 GHz and the beams from 33 to 
10 arcmin. Its noise at 4 $\mu$K at 30 GHZ to 12 $\mu$K at 100 GHz is 
significantly lower than that of MAP. The six HFI bands cover 100 to 857 GHz 
and adding them would increase the signal-to-noise of the measurement of bulk 
flows with Planck.
Because the instrument noise dipole would be added in quadrature to the 
cosmological component, 
this term is expected to be the dominant dipole noise term for MAP 
instruments. 

{4. \bf Results and strategies for measurement}.\\
In order to apply this method we will take 
the available all-sky catalogs of imaged X-ray clusters and  
compute the dipole of the CMB temperature field at the cluster locations in 
the expectation that the noise terms will integrate down uncovering the bulk 
motion contribution to the dipole. 

To measure the large-scale bulk flows by this method we need 
information on the location of $\sim$(100-300) X-ray clusters with reasonably 
measured $\tau$. A 
new all-sky catalog of imaged X-ray clusters should be available soon 
(B\"ohringer et al. 1999) from the ROSAT observations. The current catalog, 
known as BCS (Bright Cluster Sample), is $\simeq$90\% complete in the 
Northern hemisphere and $|b_{\rm Gal}|$$\geq$20 deg and contains 199 
clusters with flux $\geq$4.45$\times$$10^{-12}$erg/cm$^2$/sec (Ebeling et 
al. 1997). 
With the XLF parameters from Ebeling et al. (1997) a sphere of 
radius 100$h^{-1}$Mpc would contain $\sim$400 X-ray clusters with 
$L_{\rm X,bol}$$\geq$$10^{42} h^{-2}$erg/sec (or $10^{-3}L_*$), $\sim$200 
clusters with $L_X$$\geq$$2.5$$\times$$10^{-3}L_*$ and $\sim$100 clusters 
with $L_X$$\geq$5$\times$$10^{-3}L_*$. For flux-limited X-ray catalogs the 
numbers are similar: for flux limit of $10^{-12}$erg/cm$^2$/sec in the 
[0.1-2.4] KeV band of ROSAT, which is the flux limit of the NORA and 
REFLEX catalogs (Guzzo 2000), the number of clusters within 100$h^{-1}$Mpc 
would be $\simeq$200 or $\simeq$120 for  
$F$$\geq$2$\times$$10^{-12}$erg/cm$^2$/sec. 

ROSAT has already completed a Southern hemisphere catalog of X-ray 
clusters, ROSAT-ESO Flux-Limited X-ray cluster survey or REFLEX, (B\"ohringer 
et 
al. 1999, Guzzo et al. 2000). The flux limit in the [0.1-2.4] KeV ROSAT band 
of the REFLEX survey is $\sim$1.5$\times$$10^{-12}$erg/cm$^2$/sec corresponding 
to $L_X($2-10KeV)$\simeq$1.7$\times$$10^{42}h^{-2}$erg/sec at the distance of 
100$h^{-1}$Mpc. (Such clusters have $\tau$$\sim$9$\times$$10^{-4}$ and 
are quite numerous even at that depth avoiding the problem of significant  
shot-noise from the rare and very high-$\tau$ clusters). The Northern 
hemisphere 
catalog of the ROSAT X-ray clusters 
(NORA) should be completed shortly (Guzzo 
2000). Altogether this would result in approximately 1,500 X-ray clusters down 
to the limiting flux $F$$\sim$1$\times$$10^{-12}$ erg/cm$^2$/sec in the 
[0.1--2.4] KeV 
band; out of these $\sim$300 would lie within $\sim$100$h^{-1}$Mpc. In 
addition, there are already 50 X-ray clusters within $\sim$100$h^{-1}$Mpc from 
ASCA searches which 
have both central temperature and electron density profile measured 
(Baumgartner et al 2000); this number is expected to more than double within 
the next year or two (Mushotzky, private communication). The ROSAT clusters 
can further be imaged very efficiently with XMM in order to obtain the 
necessary catalog of X-ray clusters by the time the MAP mission is completed. 

Our ability to determine a bulk flow of 
amplitude $V_{\rm bulk}$ is limited by the instrumental noise term. The 
signal-to-noise ratio, $\chi$, in such 
a measurement out to the depth $D$ would be given by:
\begin{equation}
\chi^2 = \sum_i \frac{\int_0^D C_{1,kin} dN_{\rm cluster} }{r_i^2} = 
T_0^2 
\frac{V_{\rm bulk}^2}{c^2} N_{\rm cluster} 
\left[ \frac{3}{ D^3} \sum_i 
\frac{\int_0^D \langle \tau_0 \rangle^2 
\Psi_i^2(D) D^2 dD}{r_i^2} \right]
\label{chi}
\end{equation}
The integral on the 
right-hand-side of eq. (\ref{chi}) accounts for the fact that the number of 
clusters contributing to the reduction in the dipole from the instrument noise 
in a thin shell $[D;D+dD]$, where the beam dilutes the central optical depth 
by $\Psi_i(D)$, is $dN_{\rm cluster}$=$4\pi [\int_{L_0}^\infty n(L_X)dL_X] D^2 
dD$. For clusters selected from an absolute luminosity limited catalogs the 
term 
$\langle \tau_0 \rangle^2$ in the integral on the RHS of eq.(\ref{chi}) would 
be 
independent of $D$; for flux-limited catalog the dependence on $D$ would come 
via $L_0$$\propto$$D^2$. In computing $\chi$ we assumed the mean 
shape of $\Psi(D)$ plotted in Fig. 1. 

If we want to determine 
just the amplitude of the bulk flow, we need the 
amplitude of the dipole SZ kinematic component. If we are to determine the 
direction of the flow, we need to measure all three dipole components: 
$a_{1,1}, a_{1,0}, a_{1,-1}$. Thus to determine the amplitude of the 
bulk velocity within a given $D$ at the 95\% c.l. we need 
$\chi^2$=3.84; if we want to resolve all three 
components of the dipole at the 95\% c.l., we need $\chi^2$=7.81. 
The error bar on the dipole components translates into
an uncertainty on the measurement of the three components
of the velocity flow, i.e., on the bulk flow direction. 
If all dipole components have been measured with the same precision,
and the velocity of the flow is measured with an uncertainty $\sigma_V$, then
the direction of the flow will be determined with an angular precision
$\delta\alpha = \sqrt{2}\sigma_V/V_{\rm bulk}$. 

Fig. 2a shows the smallest
$V_{\rm bulk}$ that can be determined at 95\% c.l. with MAP and Planck/LFI 
data at $D$=50,100,150$h^{-1}$Mpc for a catalog with fixed lower bound of 
the bolometric X-ray luminosity, $L_{0,\rm bol}$. The bulk 
velocity at $150h^{-1}$Mpc from Lauer \& Postman (1994) is also shown for 
comparison.
There is only a weak dependence in $\chi^2$ on the (fixed) lower bound on the 
X-ray luminosity of the cluster catalog. Thus this method gives a very 
realistic 
way to measure 
the possible bulk flows on large scales or to significantly constrain their 
amplitude if the latter is small. 
Fig. 2b shows $\chi^2$ vs the depth of the cluster catalog for the bulk flow of 
$V_{\rm bulk}$=600 km/sec for MAP (lower set of lines) and Planck/LFI 
parameters 
(upper set of lines). Since $\chi^2$$\propto$$ V_{\rm bulk}^2$, at the 95\% 
c.l. 
the 2-year MAP data can probe in this way bulk flows with the amplitude as low 
as 
$\simeq$250 km/sec on scale of $\sim$$100h^{-1}$Mpc; with Planck data such bulk 
flows can be probed to even much lower amplitudes and scales.
Concerning the direction,  MAP and Planck/LFI would determine
\'a la Lauer-Potsman bulk flow direction with a 95\% c.l. error of $41^o$
and $16^o$, respectively.

Finally, because the upcoming X-ray catalogs, such as the NORA/REFLEX catalog, 
extend to significantly beyond $100h^{-1}$Mpc this 
method also would probe bulk flows on much larger scales. Fig.2c shows the 
amplitude of the bulk flow that can be 
determined at 95\% c.l. vs the depth of the \underline{flux-limited} X-ray 
cluster catalog. 
Bulk flows $>$100km/sec on scale $\sim$300$h^{-1}$Mpc can be probed with 
this method with MAP data; with Planck/LFI this reduces 
to $\sim$30km/sec. If MAP operates longer than 2 years, this number would 
decrease $\propto$$t_{\rm operation}^{-1/2}$. If both LFI and HFI Planck 
channels are combined the bulk 
flows on large scales can be probed to even lower limits 
than with the Planck/LFI alone; e.g. on scale of $\sim 100 h^{-1}$Mpc the 
amplitude of the bulk flows can be determined at 95\% c.l. down to 
$\sim$30km/sec and on scale of $\sim$300$h^{-1}$Mpc 
to $\sim$15-20km/sec.

AK acknowledges the hospitality and visiting support of 
the University of Salamanca. FAB acknowledges financial 
support of the Junta de Castilla y Le\'on (project SA 19/00B) and
Ministerio de Educaci\'on y Cultura (project PB 96-1306). We thank  
Chuck Bennett, Gary Hinshaw, John Mather and Richard Mushotzky for 
fruitful discussions on CMB and X-ray parts of this work. 


\newpage

FIGURE CAPTIONS

{\bf Fig. 1}: The dilution factor $\Psi(D)$ vs the depth $D$ for two MAP 
bands. 
Solid lines are for Gaussian beam corresponding to Band 1 (22 GHz, FWHM = 0.93 
deg). Dashed lines are for Band 5 (90 GHz, FWHM = 0.21 deg). Thick lines 
correspond to the cluster profile averaged over 37 X-ray clusters as described 
in the main text. Thin lines correspond to the r.m.s. value of $\Psi(D)$ over 
the 
same sample.

{\bf Fig. 2}: (a) The amplitude of $V_{\rm bulk}$ that can be determined at 
95\% c.l. ($\chi^2=3.84$) is plotted vs the lower limit on the bolometric X-ray 
luminosity, $L_{0,\rm bol}$. Thin lines correspond to MAP and thick lines to 
Planck/LFI data parameters. Solid lines correspond to
X-ray clusters out to $D=50, 100, 150 h^{-1}$Mpc from top to bottom 
respectively. From top to bottom, dashed lines show the value 
of $V_{\rm bulk}$ when the three components of the dipole 
can be determined at the 95\% c.l. at $D=50, 100, 150 
h^{-1}$Mpc, respectively. Open square with error bar shows 
the amplitude of the bulk flow at $150h^{-1}$Mpc from Lauer \& Postman (1994). 
(b) Solid lines plot 
$\chi^2$ from eq. (\ref{chi}) vs the depth of the X-ray cluster catalog, $D$, 
with a cutoff on the X-ray luminosity of 
$L_{0}/L_*=(1,2,5)\times 10^{-3}$ from top to bottom respectively. Dotted 
lines correspond to flux-limited X-ray cluster catalog 
with the [0.1-2.4] KeV flux $\geq (1,2)\times 10^{-12}$erg/cm$^2$/sec. 
Lower set of lines is for MAP data and the upper 
set of lines is for the Planck/LFI instrument parameters. In the latter case 
over the range of the plot the two lines for the [0.1-2.4] KeV 
flux $\geq (1,2)\times 
10^{-12}$erg/cm$^2$/sec coincide with each other and with the $L_X/L_* \geq 
10^{-3}$ 
case. Thick solid horizontal line corresponds to $\chi^2=3.84$, when the 
amplitude 
of the bulk flow can be determined at 95\% confidence level; thick dashed 
horizontal line corresponds to $\chi^2=7.81$, when the three components of
the dipole can be determined at 95\% confidence level. (c) The 
amplitude of the bulk flow that can be determined at 95.4\% c.l. 
is plotted vs the depth of the flux-limited X-ray cluster catalog. Solid lines 
correspond to the MAP data, thin dotted lines to Planck/LFI and thick solid 
lines to Planck/LFI and Planck/HFI together. Two sets of lines correspond to 
X-ray flux $\geq (1,2)\times 10^{-12}$erg/cm$^2$/sec from top to bottom 
respectively. 

\newpage
\clearpage
\begin{figure}
\centering
\leavevmode
\epsfxsize=1.0
\columnwidth
\epsfbox{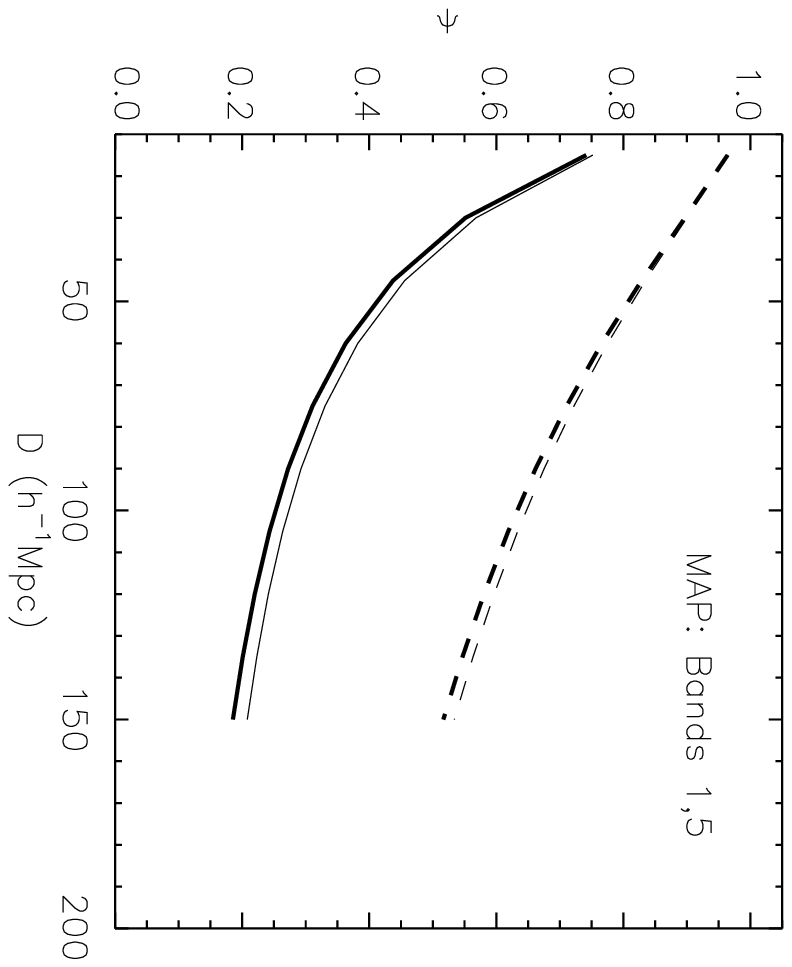}
\caption[]{ }
\label{f1}
\end{figure}

\newpage
\clearpage
\begin{figure}
\centering
\leavevmode
\epsfxsize=1.0
\columnwidth
\epsfbox{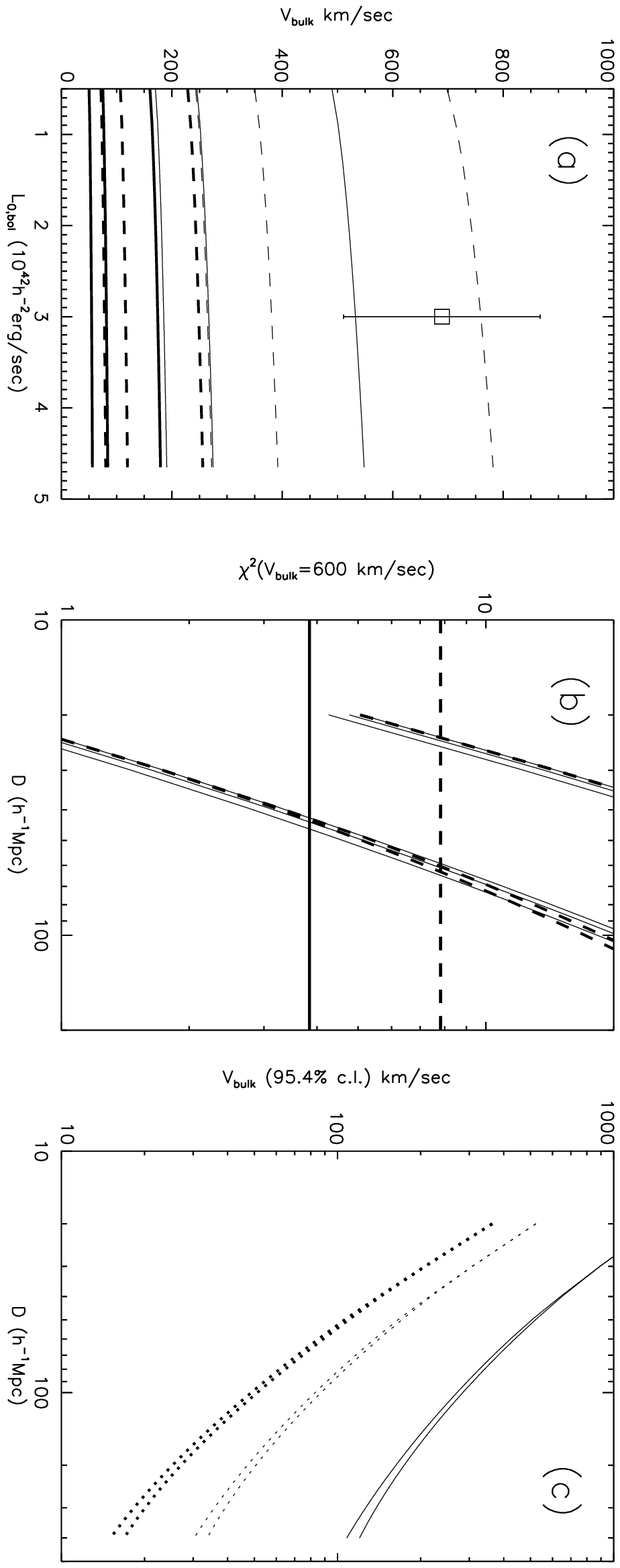}
\caption[]{ }
\label{f2}
\end{figure}

\end{document}